\documentclass{jpsj2}
\newcommand{\R}{{\mathbb R}}
\newcommand{\ga}{geometric algebra}
\newcommand{\beq}{\begin{equation}}
\newcommand{\eeq}{\end{equation}}
\newcommand{\rme}{\mathrm{e}}

\title{An Analysis of the Quantum Penny Flip Game using Geometric Algebra}

\author{
James M. \textsc{Chappell}$^{1}$\thanks{
E-mail address: james.m.chappell@adelaide.edu.au}, 
Azhar \textsc{Iqbal}$^{2,3}$, 
M. A. \textsc{Lohe},
Lorenz von \textsc{Smekal}
}

\inst{
$^{1}$Department of Physics, University of Adelaide, SA
5005, Australia.\\
$^{2}$School of Electrical \& Electronic Engineering, University of
Adelaide, SA 5005, Australia.\\
$^{3}$Centre for Advanced Mathematics \& Physics, National University
of Sciences \& Technology, Pakistan.}

\abst
{
We analyze the quantum penny flip game using \ga\
and so determine all possible unitary transformations
which enable the player $Q$ to implement a winning strategy.
Geometric algebra provides a clear visual picture
of the quantum game and its strategies, as well as providing a 
simple and direct derivation of the winning transformation,
which we demonstrate can be parametrized by two angles $\theta,\phi$.
For comparison we derive the same 
general winning strategy by conventional means using density matrices.
}

\kword{quantum game, penny flip, geometric algebra, rotor, density matrix}

\begin{document}
\maketitle


\section{Introduction}

In 1999 Meyer\cite{MeyerDavid} 
introduced the quantum version of the penny flip game, a seminal paper for 
quantum game 
theory\cite{%
Enk,Meyer2, 
Vaidman, EWL, BenjaminHayden, Johnson, Piotrowski, Du, 
ShimamuraOzdemir, OzdemirShimamura, Shimamura, 
CheonTsutsui, IchikawaTsutsui, Ozdemir, 
IqbalCheon, Ichikawa%
}.
In the classical form of this game a coin is placed heads up inside a box
so that the state of the coin is hidden from the players.
The first player $Q$ then either flips the coin or leaves it unchanged,
following which the second player $P$  also either flips the coin or not, 
and finally $Q$ flips the coin or not, after which the coin is
inspected. If the coin is heads up $Q$ wins, otherwise $P$ wins.
Classically each player has an equal chance of winning and
the optimal strategy, in order to prevent each player predicting the other's
behaviour, is to randomly flip the coin or not,
corresponding to a mixed strategy Nash equilibrium\cite{Rasmusen}.

In the quantum version of the game $P$ is restricted to classic strategies
whereas $Q$ adopts quantum strategies and so is able to
apply unitary transformations to the possible states of the coin,
which behaves like a spin half particle with a general state 
$|\psi \rangle =\alpha |0\rangle +\beta |1\rangle$,
where $|0\rangle $ and $|1\rangle $ are orthonormal states representing
heads and tails respectively, and $\alpha ,\beta $ are complex numbers.
Meyer identifies a winning strategy for $Q$ as the application of the Hadamard
transform, in which case the operation by $P$ has no effect: 
\begin{equation}
|0\rangle \overset{Q}{\longrightarrow }\frac{1}{\sqrt{2}}(|0\rangle
+|1\rangle )\overset{P}{\longrightarrow }\frac{1}{\sqrt{2}}%
(|1\rangle +|0\rangle )\overset{Q}{\longrightarrow }|0\rangle.
\end{equation}%
The final Hadamard operation
by $Q$ returns the state to the starting position, thus $Q$
wins every game using a quantum strategy.
In simple terms we can view the Hadamard operation as placing the coin ``on its
edge", which is why the flip operation of the following player has no
effect. 

Our aim in this paper is firstly to find the most general unitary 
transformations which lead to a winning strategy for $Q$, and secondly to
demonstrate that \ga\ provides a convenient formalism with which 
to find the general solution, which is
parametrized by angles $\theta,\phi$.
Our motivation in using \ga\ is ultimately to investigate quantum
mechanical correlations in strategic interactions between two
or more players of quantum games, and more generally
to exploit the analytical tools of 
game theory to better understand quantum correlations. We demonstrate in
Section \ref{dm}, however, that
for the quantum penny flip game conventional methods of analysis 
using density matrices\cite{NielsenChuang} 
are also effective in analyzing this game, and run parallel to the
\ga\ approach,  but we believe that
for $n$-player games the concise formalism of \ga\ is advantageous.

Geometric algebra\cite{GA,GA1,GA2,GA3}
is a unified mathematical formalism which simplifies the treatment of
points, lines, planes
in quantum mechanical spin half systems \cite{GA4}. 
In general, given a linear vector 
space $V$ with elements $u,v,\dots$ we may form\cite{Sz}
the tensor product $U\otimes V$ of
vector spaces $U,V$ containing elements (bivectors) $u\otimes v$.
The vector space may be extended to a vector space $\Lambda(V)$
of elements consisting of multivectors which can be
multiplied by means of the exterior (wedge) product $u\wedge v$. 
The noncommutative geometric product $uv$ of two vectors $u,v$ is defined by 
$uv=u\cdot v+u\wedge v$, which is the sum of the scalar inner product and 
the bivector
wedge product, and may be extended to the geometric product of any two 
multivectors.

Properties of the Pauli algebra have previously been developed\cite{GA4} in the 
context of \ga.  Denote by $\{\sigma_i\}$ an orthonormal 
basis in $\R^3$, then
$\sigma_i\cdot\sigma_j=\delta_{ij}$. We also have
$\sigma_i\wedge\sigma_i=0$ for each $i=1,2,3$ and so in terms of the geometric
product  we have $\sigma_i^2=\sigma_i\sigma_i=1$, and 
$\sigma_i\sigma_j=\sigma_i\wedge\sigma_j= -\sigma_j\sigma_i$ for each $i\ne j$.
Hence the basis vectors anticommute with respect to the geometric product.  
Denote by $\iota$ the trivector
\beq
\iota=\sigma_1\sigma_2\sigma_3,
\eeq
where the associative geometric product $\sigma_1\sigma_2\sigma_3$
of a bivector $\sigma_1\wedge\sigma_2$ and an
orthogonal vector $\sigma_3$ is defined by 
\[
\sigma_1\sigma_2\sigma_3=
(\sigma_1\wedge\sigma_2)\sigma_3=
\sigma_1\wedge\sigma_2\wedge\sigma_3.
\]
We have $\sigma_1\sigma_2=\sigma_1\sigma_2\sigma_3\sigma_3=\iota\sigma_3$ 
and so $\sigma_i\sigma_j=\iota\sigma_k$ for cyclic $i,j,k$.
We also find by using anticommutativity, associativity, and $\sigma_i^2=1$ that
$\iota^2=\sigma_1\sigma_2\sigma_3\sigma_1\sigma_2\sigma_3=-1$ and, furthermore,
that $\iota$ commutes with each vector $\sigma_i$. We may summarize the algebra
of the basis vectors $\{\sigma_i\}$ by the relations
\beq
\label{e3}
\sigma_i\sigma_j=\delta_{ij}+\iota\varepsilon_{ijk}\sigma_k,
\eeq
which is isomorphic to the algebra of the Pauli matrices.

We also require the following well known result in \ga. For any unit vector $u$
we can rotate a vector $v$ by an angle $\theta$ in the plane perpendicular 
to $u$ by applying a rotor $R$ defined by
\begin{equation}
\label{e5}
R=\rme^{\iota\theta u/2}=
\cos\frac{\theta}{2}+\iota u\sin\frac{\theta}{2},
\end{equation}
which acts according to $v\stackrel{R}{\longrightarrow} v'=RvR^{\dagger}$. 
$R$ is unitary in the sense that $RR^{\dagger}=R^{\dagger}R=1$, where
the $\dagger$ operation acts by 
inverting the order of terms and flipping the sign of $\iota$, and 
corresponds to the Hermitean conjugate when acting on the Pauli matrices.  
The even subalgebra of the geometric algebra of multivectors
consists of grade
zero and grade two multivectors which correspond to a spinor with four
real components.
In summary, the spinor algebra of the Pauli matrices
and the unitary matrices which rotate a polarization axis as displayed
on the Bloch sphere
may be analyzed by means of geometric algebra in three dimensions
in which vectors are operated on by a rotor\cite{GA4}.


\section{The Quantum Penny Flip Game using Geometric Algebra}

\begin{figure}[tb]
\begin{center}
\includegraphics[width=100mm]{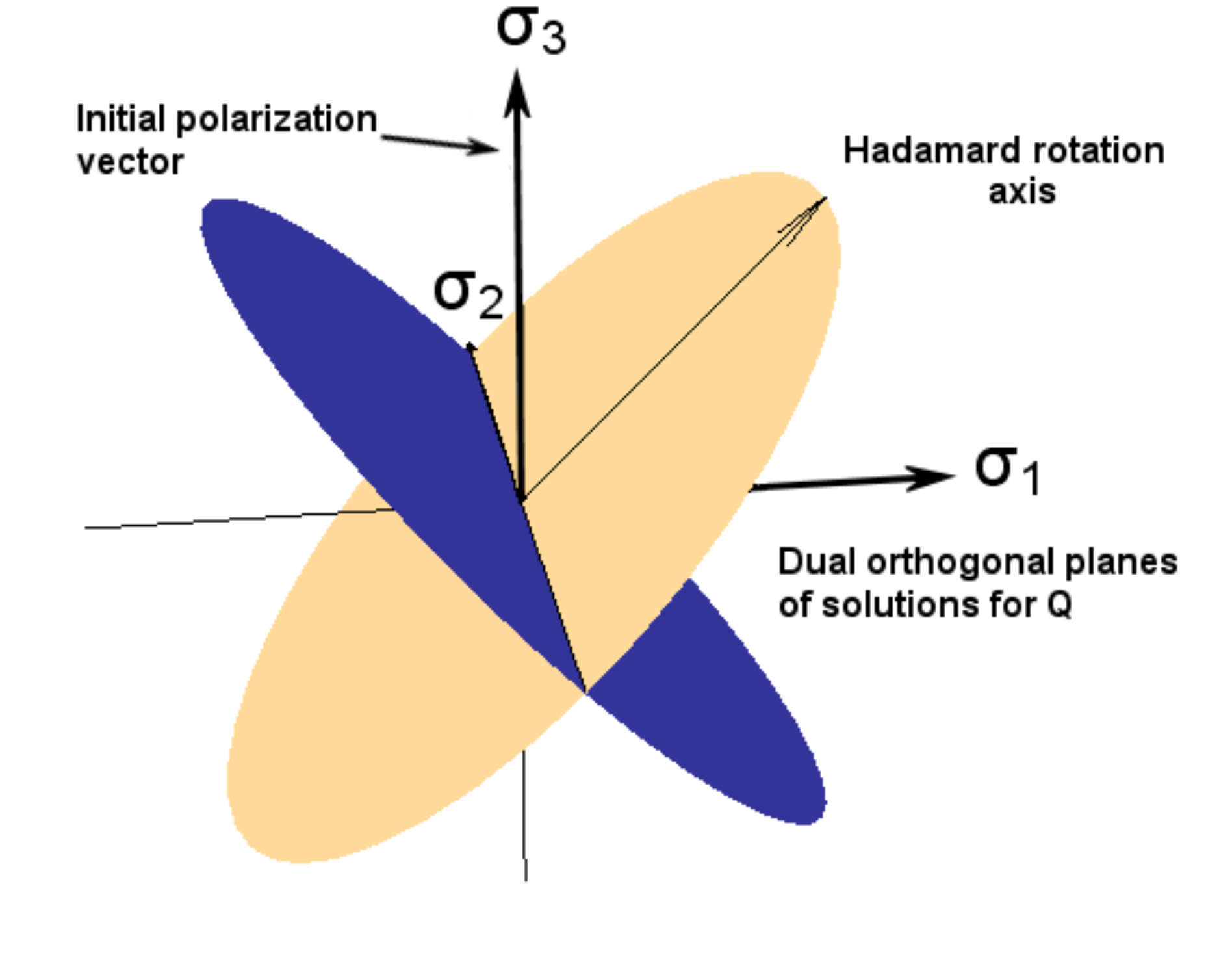}
\end{center}
\caption{
The two intersecting planes contain the unit vector $u$ which 
defines the axis of rotation about which $Q$ rotates $\sigma_3$ to
$\pm\sigma_1$, through an angle $\theta$ (color online).
}
\label{Fig1}
\end{figure}

The state of the quantum coin for heads up is $|0\rangle$
which is depicted on the Bloch sphere by the polarization vector 
pointing up on the $\sigma_3$ axis, corresponding to the initial vector
$\psi_0=\sigma_3$ as shown in Figure \ref{Fig1}.
Following operations performed by $Q,P,Q$ in turn, in which $Q$ always wins,
the final wavefunction $\psi_3$ also corresponds to the unit vector $\sigma_3$.

Suppose $Q$ first applies a general unitary transformation, represented by a
rotor (\ref{e5}), namely
$U_1=\rme^{\iota\theta u/2}$ to obtain the state
$\psi_1=U_1\psi_0U_1^{\dagger}$.
$P$ now applies the optimal classical strategy of applying a coin flip
operation $F$ with probability $p$ and no flip operation $N$ with
probability $1-p$, to obtain the mixed state 
\beq
\nonumber
\psi_2 =
pF\psi_1F^{\dagger}+(1-p)N\psi_1N^{\dagger} 
=
pFU_1\psi_{0}U_1^{\dagger}F^{\dagger}+(1-p)NU_1\psi_0U_1^{\dagger}N^{\dagger}.
\eeq
The coin flip $F$ is equivalent to the action on the spinor 
$(|0\rangle,|1\rangle)$ of the Pauli matrix $\sigma_1$ which
is isomorphic to $\sigma_1$ in \ga, so we have simply $F=\sigma_1$ and
also $N=1$.
$Q$ now applies a final unitary transformation $U_3$ which is independent
of $p$ to obtain
\beq
\label{e1}
\psi_3 
=
U_3\psi_2U_3^{\dagger} 
=
pU_3\sigma_1U_1\sigma_3U_1^{\dagger}\sigma_1U_3^{\dagger}
+(1-p)U_3U_1\sigma_3U_1^{\dagger}U_3^{\dagger}.  
\eeq
Since we assume that $Q$ always wins, i.e.\ $\psi_3=\sigma_3$ for any $p$, 
the terms in this expression must equal $p\sigma_3$ and $(1-p)\sigma_3$
respectively.  For the second term this requires 
$U_3U_1\sigma_3U_1^{\dagger}U_3^{\dagger}=\sigma_3$ 
and so $U_3U_1$ must commute with $\sigma_3$.
Hence $U_3U_1=\rme^{\iota\phi\sigma_3/2}$ for some angle 
$\phi$, i.e.\ $U_1= U_3^{\dagger}\rme^{\iota\phi\sigma_3/2}$.
On substituting into (\ref{e1}) we find 
\beq
\label{e2}
\psi_3 =
pU_3\sigma_1U_3^{\dagger}\sigma_3U_3\sigma_1U_3^{\dagger}+(1-p)\sigma_3,
\eeq
which has no explicit dependence on the angle $\phi$ which 
therefore remains arbitrary.
Evidently it is not necessary that $U_3$ be inverse to the initial
rotation $U_1$, i.e.\ the final rotation need not be about the same
axis as the initial rotation.
In order for the first term in (\ref{e2}) to equal $p\sigma_3$ we require
$U_3\sigma_1U_3^{\dagger}\sigma_3U_3\sigma_1U_3^{\dagger}=\sigma_3$,
that is,
\beq
\label{e4}
U_3\sigma_1U_3^{\dagger}\sigma_3=\sigma_3U_3\sigma _1U_3^{\dagger},
\eeq
and so $U_3\sigma _1U_3^{\dagger}$ commutes with $\sigma_3$. This
implies that $U_3\sigma _1U_3^{\dagger}$ is a multiple of $\sigma_3$,
since the rotated vector $U_3\sigma _1U_3^{\dagger}$ is
a linear combination of the basis elements, i.e.\
$U_3\sigma _1U_3^{\dagger}=c_1\sigma_1+c_2\sigma_2+c_3\sigma_3$ for 
some scalars
$c_i$, and the Pauli algebra (\ref{e3}) then
implies that (\ref{e4}) is satisfied only if $c_1=c_2=0$. Since
$U_3\sigma _1U_3^{\dagger}$ is a unit vector we also have $c_3=\pm1$.

The final state $\psi_3$ is therefore equal to $\sigma_3$, namely heads up
independent of $p$, provided 
$U_3= \rme^{\iota\phi\sigma_3/2}U_1^{\dagger}$ 
and $U_1\sigma_3U_1^{\dagger}=\pm\sigma_1$.  
Hence $Q$'s strategy is clear: by rotating the starting
vector $\sigma_3 $ to $\pm\sigma_1$, $P$'s coin flip
operation has no effect because $F\sigma_1F^{\dagger}=\sigma
_1\sigma_1\sigma_1=\sigma_1$, and so $Q$ simply then applies 
$U_3= \rme^{\iota\phi\sigma_3/2}U_1^{\dagger}$ 
to turn the coin back to heads where it started.


\section{\label{MQ}Solution for $Q$'s Winning Strategy}

By substituting for the general rotor $U_1=R$ 
as given in (\ref{e5}), and by writing the unit vector as
$u=a\sigma_1+b\sigma_2+c\sigma_3$ where the scalars $a,b,c$ 
satisfy $a^2+b^2+c^2=1$, we find that we require 
$R^{\dagger}\sigma_1=c_3\sigma_3R^{\dagger}$, specifically
\beq
\label{f4}
\left[\cos\frac{\theta}{2}-\iota\sin\frac{\theta}{2}
(a\sigma_1+b\sigma_2+c\sigma_3)\right] 
\sigma_1
=
c_3\sigma_3
\left[\cos\frac{\theta}{2}-\iota\sin \frac{\theta}{2}
(a\sigma_1+b\sigma_2+c\sigma_3)\right], 
\eeq
where $c_3^2=1$. This equation is satisfied if and only if
$a\sin\frac{\theta}{2}=c_3c\sin \frac{\theta}{2}$ and
$b\sin\frac{\theta}{2}=c_3\cos\frac{\theta}{2}$, which implies
$\sin\frac{\theta}{2}\ne0$. Hence $a=c_3c$ and $b=c_3\cot\frac{\theta}{2}$.
Since $u$ is a unit vector  we have $2a^{2}+\cot ^{2}\frac{\theta}{2}=1$
which implies $|\cot\frac{\theta}{2}|\leqslant1$, and 
hence $\theta$ can take any value such that
$\frac{\pi}{2}\leqslant|\theta|\leqslant\frac{3\pi}{2}$. 
Then
\begin{equation}
\label{e7}
a=\pm \sqrt{\frac{1}{2}-\frac{1}{2}\cot^{2}\frac{\theta}{2}}\;,
\end{equation}
together with $c=c_3a$ and $b=c_3\cot\frac{\theta}{2}$ where $c_3=\pm1$.
Thus we have the general expression
\begin{equation}
\label{e6}
U_1
=
\rme^{\iota\theta
(a\sigma_1+c_3\cot\frac{\theta}{2}\sigma_2+c_3a\sigma_3)/2},
\end{equation}
with which $Q$ rotates $\sigma_3$ to $\pm\sigma_1$ about the axis defined by
$u$ through the angle $\theta$,
for any $\theta$ in the specified range. The unit vector 
$u=a\sigma_1+b\sigma_2+c\sigma_3$ lies in one of the two intersecting 
planes defined by $|a|=|c|$, as shown in Figure \ref{Fig1}.
Denote the angle between $u$ and $\sigma_3$ by $\psi$
then $\cos\psi=\frac{1}{2}(\sigma_3 u+u\sigma_3)=c=\pm a$ which implies
$-1/\sqrt{2}\leqslant\cos\psi\leqslant1/\sqrt{2}$, showing that $u$ is tilted
with respect to $\sigma_3$ at an angle $\psi$ in the range
$\pi/4\leqslant|\psi|\leqslant3\pi/4$.

The choice of sign for $a$ in (\ref{e7}) can in effect be altered
by replacing $\theta\to-\theta$ in (\ref{e6}), and the
sign $c_3=\pm1$ can be reversed by replacing 
$\sigma_2\to-\sigma_2, \sigma_3\to-\sigma_3$, which leaves the Pauli
algebra (\ref{e3}) invariant, and may be implemented
by rotating the system about the $\sigma_1$ axis through $\pi$ by 
means of the rotor 
$S=\rme^{\iota\pi\sigma_1/2}=\iota\sigma_1$. The final move by $Q$
is the rotation
$U_3= \rme^{\iota\phi\sigma_3/2}U_1^{\dagger}$ 
which depends on two parameters $\theta,\phi$, where
$\rme^{\iota\phi\sigma_3/2}$  performs a rotation about the
$\sigma_3$ axis leaving $\sigma_3$ unchanged.

We recover Meyer's solution by choosing $\theta=\pi,\phi=0$ together
with appropriate signs, to obtain
$U_1= \rme^{\iota\frac{\pi}{2}\left(\sigma_1+\sigma_3\right)/\sqrt{2}}
= U_3^{\dagger}$,
which performs a rotation of $\theta=\pi$ about the line defined 
by the vector $(\sigma_1+\sigma_3)/\sqrt{2}$, 
and so reproduces the Hadamard transform 
which rotates the polarization vector onto the $\sigma_1$ axis as shown in
Figure \ref{Fig1}. 


\section{\label{dm}Analysis using Density Matrices}

The analysis of the quantum penny flip game using geometric algebra can
be reproduced by means of density matrices and unitary 
transformations\cite{NielsenChuang}.
We may write any $2\times2$ unitary
matrix in the form $U=\rme^{i A}$ where the Hermitean matrix $A$ can be 
expanded in terms of the Pauli matrices $\sigma_i$ and the identity matrix 
$I_2$ according to
$A=\alpha(a\sigma _{1}+b\sigma_{2}+c\sigma _{3})+\beta I_2$ where the 
scalars $a,b,c$ are normalized such that $a^{2}+b^{2}+c^{2}=1$, and 
where $\alpha,\beta$ are fixed angles.  
If we define
$\theta=2\alpha$ and also the $2\times2$ matrix
$\widehat{u}=a\sigma _{1}+b\sigma_{2}+c\sigma _{3}$ (which satisfies
$\widehat{u}^2=I_2$), then 
\beq
\label{f2}
U= \rme^{i\widehat{u}\theta/2}\rme^{i\beta}
=
\left(I_2\cos\frac{\theta }{2}+i\widehat{u}\sin\frac{\theta}{2}\right)
\rme^{i\beta},
\eeq
which compares with the expression (\ref{e5}) for the rotor $R$. We emphasize,
however, that in (\ref{e5}) the element $\iota$  is a tri-vector and
$u$ denotes a unit vector which is a linear combination of basis
vectors $\sigma_i$.

If we denote the starting state by $|0\rangle=
\left(\begin{smallmatrix} 1\\0 \end{smallmatrix}\right)$,
then the first move by $Q$ 
is to apply a general unitary transformation $U_1$
on the starting density matrix $\rho _{0}=|0\rangle \langle 0|$,
which therefore evolves to $\rho _{1}=U_{1}\rho _{0}U_{1}^{\dagger }$.
$P$ now applies the optimal classical strategy of applying a coin flip
operation $F=\sigma _{1}$ with probability $p$ and the no flip operation 
$N=I_2$ with probability $1-p$ producing 
\[
\rho _{2} =pF\rho _{1}F^{\dagger }+(1-p)N\rho _{1}N^{\dagger } 
=p\sigma _{1}U_{1}\rho _{0}U_{1}^{\dagger }\sigma _{1}+(1-p)U_{1}\rho
_{0}U_{1}^{\dagger }. 
\]
$Q$ applies a final unitary transformation $U_{3}$ which is independent
of $p$ to obtain 
\beq
\label{f1}
\rho _{3}=U_{3}\rho _{2}U_{3}^{\dagger }
=
pU_{3}\sigma _{1}U_{1}\rho_{0}U_{1}^{\dagger }\sigma _{1}U_{3}^{\dagger }
+
(1-p)U_{3}U_{1}\rho_{0}U_{1}^{\dagger }U_{3}^{\dagger }, 
\eeq
which is a matrix equation which can be compared with the geometric
algebraic expression (\ref{e1}), in which 
the unit vector $\sigma _{3}$ replaces the
initial density matrix $\rho _{0}$ 
and $U_3$ implements a quaternion rotation of $\sigma _{3}$.
Since we assume that $Q$ always wins, i.e.\ that $\rho _{3}=|0\rangle \langle
0|$ for any $p$, the terms in the expression (\ref{f1})
must equal $p|0\rangle \langle
0|$ and $(1-p)|0\rangle \langle 0|$ respectively, which for the second term 
requires 
$U_3U_1|0\rangle\langle 0|U_1^{\dagger }U_3^{\dagger}=|0\rangle\langle 0|$,
where $U=U_3U_1$ is a unitary matrix.

The general solution of the matrix equation 
$U|0\rangle\langle 0|U^{\dagger}=|0\rangle\langle 0|$ for any $2\times2$ unitary
matrix $U$, where
$|0\rangle\langle 0|
=\left(\begin{smallmatrix}1&0\\0&0\end{smallmatrix}\right)
=\frac{1}{2}(\sigma_3+I_2)$,
may be found by parametrizing $U$ as above or, more directly,
observing that this equation is equivalent to $[U,\sigma_3]=0$. The solution is
$U=\rme^{i\beta}\rme^{i\sigma_3 \phi/2}$
for angles $\beta,\phi$.

Hence we have $U_3U_1=\rme^{i\beta}\rme^{i\sigma_3\phi/2}$ and on substituting 
$U_1|0\rangle\langle 0|U_1^{\dagger }=U_3^{\dagger}|0\rangle\langle 0|U_3$ 
into Eq.~(\ref{f1}) we find 
\begin{equation}
\rho _{3}=pU_{3}\sigma _{1}U_{3}^{\dagger }|0\rangle \langle 0|U_{3}\sigma
_{1}U_{3}^{\dagger }+(1-p)|0\rangle \langle 0|,  \label{eq:rho3firstA}
\end{equation}%
which has no explicit dependence on the angle $\phi $ which therefore
remains arbitrary. In order that the first term in Eq. (\ref{eq:rho3firstA})
equal $p|0\rangle \langle 0|$ we require 
\begin{equation}
U_3\sigma_1U_3^{\dagger}|0\rangle\langle 0|U_3\sigma_1U_3^{\dagger}
=|0\rangle\langle0|,
\end{equation}%
where $U=U_3\sigma_1U_3^{\dagger}$ is unitary.
As discussed above, this matrix equation is equivalent to
$[U,\sigma_3]=0$ which implies that $U$ is a linear combination of $I_2$ and
$\sigma_3$. We also have $U^2=I_2$ which implies, since $U\ne\pm I_2$, that
$U=\pm\sigma_3=U_3\sigma_1U_3^{\dagger}$.

Thus the final state is heads up independent of $p$, provided 
$U_3=\rme^{i\beta}\rme^{i\sigma_3\phi/2}U_1^{\dagger}$ 
and $U_1\sigma_3U_1^{\dagger}=\pm\sigma_1$. The phase angle $\beta$ can be set 
to zero without loss of generality. By substituting for the general
unitary transformation $U_1=U$ as given by Eq.~(\ref{f2}) we
require $U^{\dagger}\sigma_1=\pm\sigma_3U^{\dagger}$, 
specifically: 
\begin{equation}
\left[I_2\cos\frac{\theta}{2}-i\sin\frac{\theta}{2}
(a\sigma_1+b\sigma_2+c\sigma_3)\right]\sigma _1
=
\pm\sigma_3\left[I_2\cos\frac{\theta}{2}-i\sin\frac{\theta}{2}
(a\sigma_1+b\sigma_2+c\sigma_3)\right],
\end{equation}
which compares with the isomorphic Eq.~(\ref{f4}) 
derived using geometric algebra, and which 
therefore has the solution Eq.~(\ref{e6}) 
in which $ \sigma_1,\sigma_2, \sigma_3 $ now refer to Pauli matrices,
instead of unit vectors.

Evidently this derivation of the general solution closely parallels that
using \ga, which uses quaternion rotations of vectors in real 3-space, with
the formalism defined in terms of unit vectors 
$\sigma_1,\sigma_2,\sigma_3$, whereas
the density matrix formalism uses Dirac's bra-ket notation, density
matrices and complex matrices for $SU(2)$ rotations.
Geometric algebra has the advantage of avoiding global phase
factors $\rme^{i\beta}$ and also permits a geometric picture as
shown in Figure \ref{Fig1}, which is 
hidden in the density matrix formalism.



\section{Conclusion}

We have determined unitary transformations,
parametrized by angles $\theta,\phi$, which enable $Q$
to implement a foolproof winning strategy for the 
quantum penny flip game. These transformations are derived using both the 
formalism of \ga, which facilitates a geometric approach, and also 
density matrices. 
The matrix condition given by Meyer\cite{MeyerDavid}
for the general solution is in effect parametrized and solved by this means.
Geometric algebra in general has the significant benefit of an 
intuitive understanding and offers better insight
into quantum games and, for the quantum penny flip game,
allows an analysis using operations in 3-space with real
coordinates, thus permitting a visualization that is helpful in 
determining $Q$'s winning strategy.
A natural extension of the present work (in progress) is to 
apply \ga\ to $n$-player
quantum games, in which all players perform local
quantum mechanical actions on entangled states, with the 
outcome determined by measurement of the final state.


\end{document}